\documentclass[twocolumn,prb,amssymb,amsmath,superscriptaddress,showpac]{revtex4}
\usepackage{graphicx}
\input{ulem.sty}

\begin{document}

\title{Diffusion conductivity and weak localization in two-dimensional
structures with electrostatically induced random antidots array}

\author{G.~M.~Minkov}
\author{A.~A.~Sherstobitov}
\affiliation{Institute of Metal Physics, RAS, 620041 Ekaterinburg,
Russia}
\author{A.~V.~Germanenko}
\author{O.~E.~Rut}
\affiliation{Institute of Physics and Applied Mathematics, Ural
State University, 620083 Ekaterinburg, Russia}

\date{\today}

\begin{abstract}
Results of experimental study of the weak localization phenomenon in
2D system with artificial inhomogeneity of potential relief are
presented. It is shown that the shape of the magnetoconductivity
curve is determined by the statistics of closed paths. The area
distribution function of closed paths has been obtained using the
Fourier transformation of the magnetoconductivity curves taken at
different temperatures. The experimental results are found in a
qualitative  agreement with the results of computer simulation.
\end{abstract}

\maketitle

The transport properties of the semiconductor two-dimensional (2D)
structures with antidots array were intensively studied both
experimentally\cite{exper,exper0,exper1,exper2,exper3,exper4,exper5,exper6,exper7,exper8,exper9}
and theoretically\cite{Theory,Theory0,Theory1,Theory2,Theory3} in
last 10--15 years. The main concern was with the investigations of
the ballistic systems, in which $l>d,\,D$, where $l$ is the mean
free path, $d$ and  $D$ are the size of antidots and period of the
antidots array, respectively. The rich  diversity of the transport
phenomena, such as commensurable oscillations, the peculiarities due
to the trajectories rolling along the array of antidots was
observed. The antidots array structures in the diffusion regime ($l
<d,D$) were  not studied essentially. Such structures are
interesting in some aspects. First of all, the quantum corrections
to the conductivity due to weak localization (WL) and
electron-electron ({\it e-e}) interaction have to reveal the
specific features when the phase breaking length,
$L_\phi=\sqrt{{\cal D}\tau_\phi}$, where ${\cal D}$ is the diffusion
coefficient and $\tau_\phi$ is phase breaking time, or the
temperature length, $L_T=\sqrt{{\cal D}/T}$, become larger than
$d,D$ at decreasing temperature (hereafter we set $k_B=1$,
$\hbar=1$). Secondly, the large enough negative gate voltage has to
deplete the channels between the antidots (the channel width $w_0$
is about $D-d$) and as a result to lead to crossover to the hopping
conductivity. Besides, from the transport properties standpoint the
antidots arrays are the fine model of the granular media. In
contrast to the granular metallic film, the parameters of the
``granules'' and ``barriers'' are reliably known and can be changed
continuously within wide range.

In this paper we report the results of the experimental study of the
weak localization correction to  the conductivity in the structure
with the random array of antidots. We show that the change of the
magnetoresistance at arising of the antidots and increase of the
antidots size results from the change of  statistics of closed
paths. Namely, the contribution of the trajectories with the large
enclosed area is strongly suppressed. The experimental area
distribution function is in a reasonable agreement with the function
obtained from the computer simulation.

\begin{figure}
\includegraphics[width=\linewidth,clip=true]{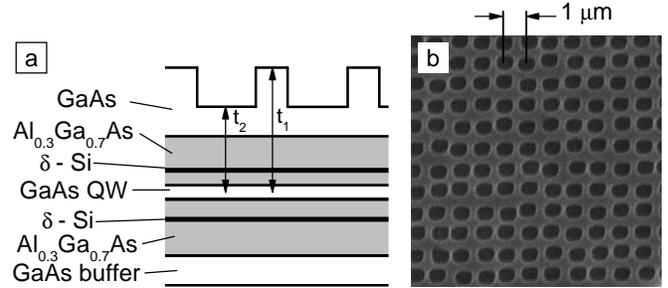}
\caption{ \label{F0} The sketch (a) and electron microscope image
(b) of the structure investigated. }
\end{figure}

The random antidots array was made on the single quantum well
heterostructure with the electron density $n=1.5\times
10^{12}$~cm$^{-2}$ and mobility $\mu=19000$~cm$^2$/(V\,s) grown by
the molecular beam epitaxy.  It consists of a 250 nm-thick undoped
GaAs buffer layer grown on semiinsulator GaAs, a 50 nm
Al$_{0.3}$Ga$_{0.7}$As barrier, Si $\delta$ layer, a 6 nm spacer of
undoped Al$_{0.3}$Ga$_{0.7}$As, a 8 nm GaAs well, a 6 nm spacer of
undoped Al$_{0.3}$Ga$_{0.7}$As, a Si $\delta$ layer, a 50 nm
Al$_{0.3}$Ga$_{0.7}$As barrier, and $150$~nm  cap layer of undoped
GaAs [Fig.~\ref{F0}(a)]. The samples were etched into standard Hall
bars. The holes in the cap layer [see Fig.~\ref{F0}(b)] were
fabricated with the use of electron beam lithography and wet
etching. Their depth measured by atomic force microscope (AFM)
consists of about $85$~nm. Note that this depth is less than the cap
layer thickness before etching. The holes of $0.7$-$\mu$m-diameter
were shifted randomly on the value $\approx 0.1$~$\mu$m from the
sites of the square lattice with the period of about $1$~$\mu$m.
This shift destroys  all oscillations in the galvanomagnetic effects
resulted from the commensurability between the cyclotron orbits and
the lattice period and from the trajectories rolling along the array
of antidots. After etching an Al gate electrode was deposited by
thermal evaporation onto the cap layer.

\begin{figure}
\includegraphics[width=\linewidth,clip=true]{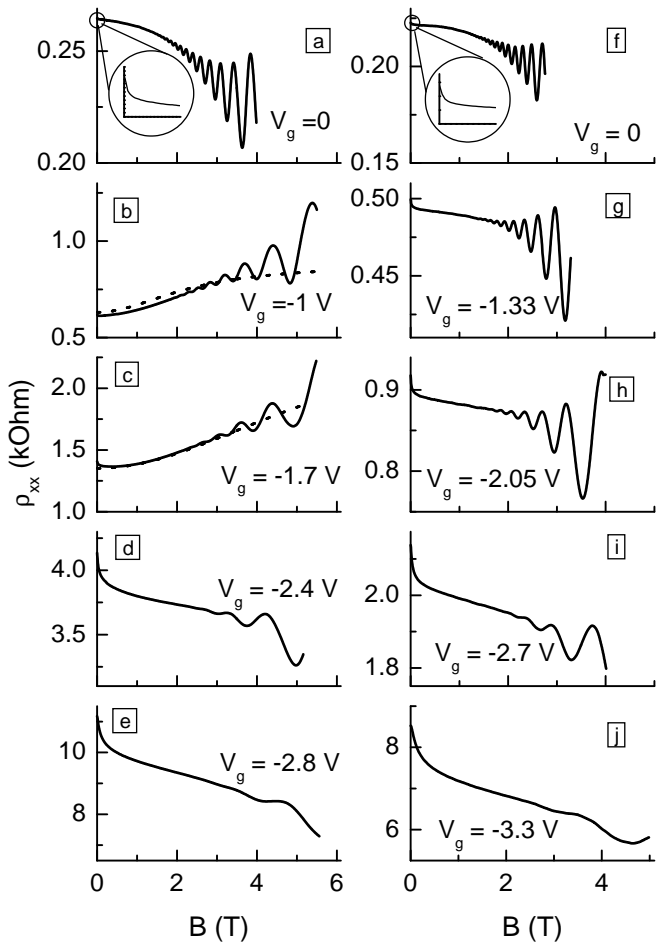}
\caption{\label{F1} The magnetic field dependences of $\rho_{xx}$
for pattern (left column) and unpattern (right column) structures
for $T=4.2$~K. The dotted lines in (b) and (c) are calculated from
the standard classical formula for two types of carriers with the
electron densities found from Eqs.~(\ref{eq10}) and (\ref{eq100}),
and $\mu_1=12600$~cm$^2$/(V\,s), $\mu_2=3400$~cm$^2$/(V\,s) for
$V_g=-1$~V, and $\mu_1=8600$~cm$^2$/(V\,s),
$\mu_2=1300$~cm$^2$/(V\,s) for $V_g=-2.4$~V that corresponds to the
model described in the text.}
\end{figure}

The magnetic field dependences of the longitudinal resistance
($\rho_{xx}$)  for the pattern structure  together with that for the
unpattern one for some gate voltages are shown in Fig.~\ref{F1}. It
is seen that they are rather complicated. Following the sharp
magnetoresistivity in low magnetic field, which results from the
suppression of the interference quantum correction, the relatively
smooth  negative or positive magnetoresistivity against the
background of the Shubnikov-de Haas (SdH) oscillations is observed.

Let us first consider the range of high magnetic field. At $V_g=0$
the magnetoresistance in the pattern and unpattern structures are
very close to each other  [see Figs.~\ref{F1}(a) and \ref{F1}(f)].
In both cases the paraboliclike negative magnetoresistance resulting
from the {\it e-e} interaction correction (for more detail see
Ref.~\onlinecite{our1}) and the SdH oscillations of close frequency
are observed. Such the behavior  is not surprising since the
difference in local electron densities in quantum well under the
holes and out of them, $\delta n$,  is relatively small at $V_g =0$.
It can be estimated as $\delta n\approx (C_2-C_1)V_s/|e|$, where
$V_s$ is the surface potential, $C_{1,2}=\varepsilon\varepsilon_0/
t_{1,2}$ is the local capacity, $t_{1,2}$ is the distance between 2D
gas and gate electrode in different locations of the structure [see
Fig.~\ref{F0}(a)], $\varepsilon_0$ is dielectric constant of free
space. With the use of $V_s\simeq 0.7$~V, $t_1=85$~nm, $t_2=125$~nm,
and $\varepsilon=12.5$ we obtain $\delta n/n\approx 0.1$. However,
$\delta n$ strongly increases with the lowering gate voltage that
leads to the positive magnetoresistance evident in the pattern
sample within the gate voltage range from $-1$~V to $-1.8$~V. The
rise of the positive magnetoresistance is a sequence of the
comparable contributions to the conductivity from the regions under
the holes and out of them. For the first approximation the transport
in such inhomogeneous media can be considered as determined by two
types of carriers with the different mobility and
density.\cite{footn1} The dotted curves in Figs.~\ref{F1}(b) and
\ref{F1}(c) are the magnetoresistivity calculated in the framework
of this simple model with parameters determined below. At
$V_g<-1.8$~V, the negative magnetoresistance is restored because the
dielectric holes in 2D gas are formed and the conductivity of the
structure is determined by the channels between the antidots.

\begin{figure}
\includegraphics[width=\linewidth,clip=true]{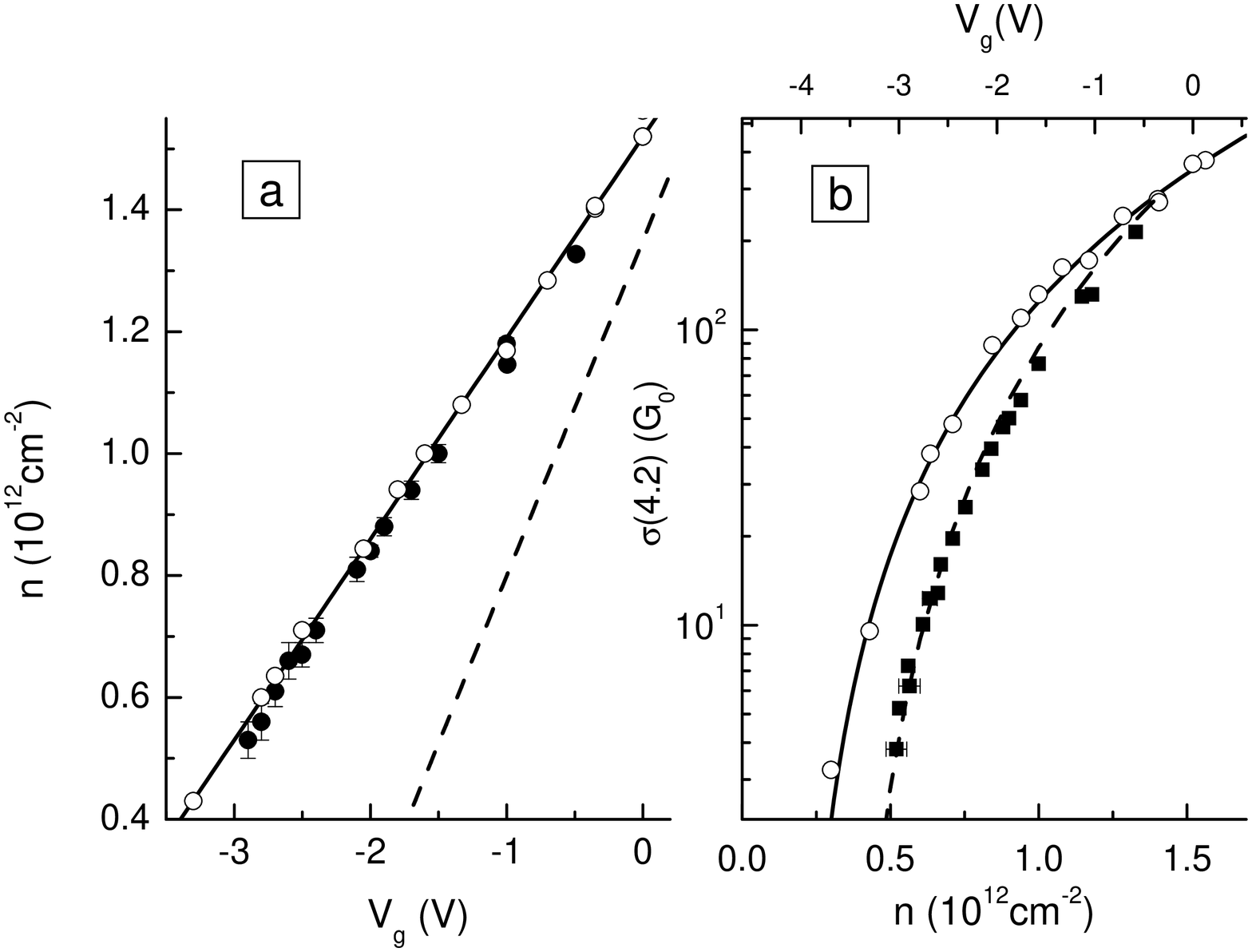}
\caption{\label{F2} (a) The gate voltage dependence of the electron
density found from the SdH oscillations for the pattern (solid
symbols) and unpattern (open symbols) structure. Solid and dashed
lines are drawn according to Eqs.~(\ref{eq10}) and (\ref{eq100}),
respectively. (b) The electron density dependences of the
conductivity measured at $T=4.2$~K for the pattern (solid symbols)
and unpattern (open symbols).}
\end{figure}

As figure~\ref{F1} shows  the SdH oscillations in the pattern
structure are observed down to $V_g=-2.8$~V.  The periods of the
oscillations in the pattern and unpattern structures at given $V_g$
are  close to each other [see Fig.~\ref{F2}(a)] and  the gate
voltage dependence of the electron density calculated from the
oscillations is well described by the expression
\begin{equation}
\label{eq10} n_1(V_g)=n(V_g)=(1.52+0.33\,V_g)\times 10^{12},\,
\text{cm}^{-2}.
\end{equation}
The facts that the oscillations of only one period are observed in
the pattern sample and the dependences $n_1(V_g)$ and $n(V_g)$ are
practically the same mean that only the areas of 2D gas located out
of the holes contribute to the SdH oscillations. The areas under
various holes have probably different electron density due to the
different depths of the holes and, therefore, the corresponding
oscillations are very broadened. The $V_g$ dependence of the
electron density under the holes $n_2(V_g)$ can be obtained from the
geometric consideration. With the use of the local cap layer
thickness $t_2=125$~nm we have
\begin{equation}
\label{eq100} n_2(V_g)=(1.35+0.55\,V_g)\times 10^{12},\,
\text{cm}^{-2}.
\end{equation}
In Fig.~\ref{F2}(a) this dependence is shown by dashed line.

Thus analyzing the high-field magnetoresistance we reason that: (i)
at $V_g<-1.8$~V the conductivity is mainly determined by the
channels; (ii)  the electron density  out of the dielectric holes
remains more or less homogeneous in this $V_g$ range.

\begin{figure}
\includegraphics[width=\linewidth,clip=true]{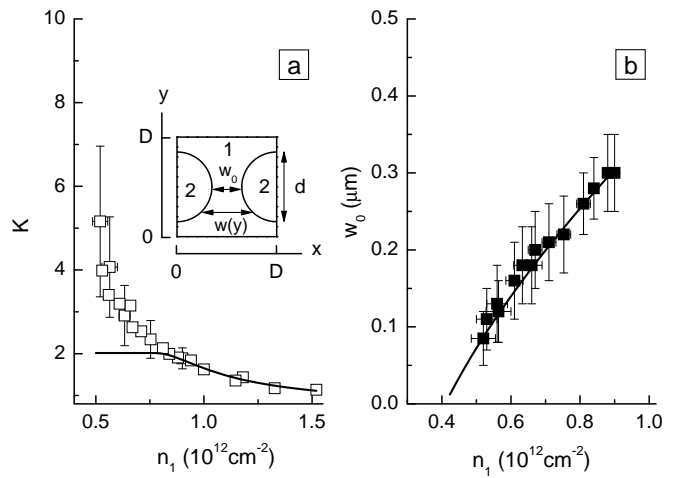}
\caption {\label{F3} (a) The $K$-vs-$n_1$ dependence obtained
experimentally (symbols) and calculated from Eq.~(\ref{eq1}) (lines)
for $d=0.7$~$\mu$m and $D=1$~$\mu$m. Insert shows one period of the
antidots array. (b) The $n_1$ dependence of the channel width $w_0$
obtained from the experimental values of $K$ at low $V_g$ with the
help of Eq.~(\ref{eq1}). The line is provided as a guide for the
eye.}
\end{figure}

Let us compare the behavior of the conductivity for pattern
($\sigma_{patt}$) and unpattern ($\sigma_{unpatt}$) samples at
$B=0$.  The values of $\sigma_{patt}$ and $\sigma_{unpatt}$ measured
at $T=4.2$~K are plotted against  the electron density found from
the SdH oscillations in Fig~\ref{F2}(b). It is seen that the
conductivity of the pattern structure significantly steeper falls
down with decreasing $n$ than that of unpattern sample.
Qualitatively such a behavior is transparent. This is because that
the 2D gas under the holes in the pattern sample is depleted faster
with lowering $V_g$ than that out of them due to thinner cap layer
in these locations. If this is the case we can obtain the
geometrical parameters of the conducting areas knowing the
experimental value of the ratio $\sigma_{unpatt}/\sigma_{patt}$
referred below as $K$. Using the data obtained for the unpattern
structure we obtain the mean free path, $l\simeq 0.03\ldots
0.3$~$\mu$m depending on the gate voltage,  being less than the
characteristic scales of the holes and channels, $\sim 0.5$~$\mu$m
[see Fig.~\ref{F0}(b)]. Therefore, one can deal with the local
conductivity. If one additionally neglects the randomness in the
antidots position, we can write out the following approximate
expression for the conductivity of the pattern structure:
\begin{equation}
\sigma_{patt}\simeq \left\{\int_0^{D}\frac{dy}{\sigma_1  w(y)
+\sigma_2 [D-w(y)]}\right\}^{-1}. \label{eq1}
\end{equation}
where $\sigma_1$ and $\sigma_2$ stand for the local conductivity of
the 2D gas out of and under the holes (in the insert of
Fig.~\ref{F3}(a) these areas labeled as 1 and 2, respectively),
$w(y)$ is the $y$-dependent width of the area 1. In what follows we
suppose the area 2 being round in the shape. The above equation
gives the result, which coincides with the exact solution with the
accuracy better than $20$ percent when $w_0/D>0.1$, where
$w_0=w(D/2)$. In order to calculate the dependence $K(V_g)$ it is
natural to suppose that the local conductivity $\sigma_1$ and
$\sigma_2$ is fully determined by the local electron density $n_1$
and $n_2$, respectively, and the $\sigma_i$-vs-$n_i$ behavior is
just the same as that for the unpattern sample $\sigma_1(n_1)$,
$\sigma_2(n_2)=\sigma_{unpatt}(n)$ [shown by open symbols in
Fig.~\ref{F2}(b)].

In Fig.~\ref{F3}(a) we present the $K$-vs-$n_1$ dependences as they
have been obtained experimentally and calculated from
Eq.~(\ref{eq1}) with $d=0.7$~$\mu$m obtained from AFM. It is seen
that the above simple model well describes the experimental results
down to $n_1\simeq 8.5\times 10^{11}$~cm$^{-2}$ that corresponds to
$V_g\simeq -2$~V. The reason for the discrepancy which is clearly
evident at lower electron density is transparent. For the fixed $d$
value, the saturation of the calculated $K$-vs-$n_1$ dependence at
$n_1< 8.5\times 10^{11}$~cm$^{-2}$ ($V_g\lesssim -2$~V) results from
the fact that $\sigma_2$ becomes much less than $\sigma_1$. The
enhance of $K$ obtained experimentally for these gate voltages is
sequence of the depletion of the area outside the antidots, i.e., of
an increase of the antidots size $d$ and decrease of the channel
width $w_0=D-d$. Thus knowing the experimental  value of $K$ and
using Eq.~(\ref{eq1}) we are able to find the channel width $w_0$
when $n_1\lesssim 8.5\times 10^{11}$~cm$^{-2}$ . The results are
depicted in Fig.~\ref{F3}(b). It is seen that the separation between
antidots $w_0$ decreases with $n_1$ decrease. Extrapolating the
$w_0$-vs-$n_1$ plot to $w_0=0$ one obtains that the antidots close
when $n_1\simeq 4\times 10^{11}$~cm$^{-2}$ [$V_g=-(3.4\ldots
3.2)$~V].

Thus, the gate voltage dependence of the conductivity and the high
magnetic field magnetoresistance are reasonably described within the
following simple model. At $V_g\simeq (0\ldots -2)$~V the
conductivity is determined both by the areas under holes and out of
the them.  These areas are characterized by the different electron
density, which determines the local conductivity. At $V_g\simeq
-(2\ldots 3)$~V the antidots are formed and the conductivity of
structure is determined by the channels between the antidots with
local conductivity $\sigma_1$ equal to the conductivity $\sigma$ of
the unpattern sample at the same electron density. Finally, at
$V_g=-(3.4\ldots 3.2)$~V the channels are collapsed and most likely
the crossover to the hopping conductivity should occur.

\begin{figure}
\includegraphics[width=\linewidth,clip=true]{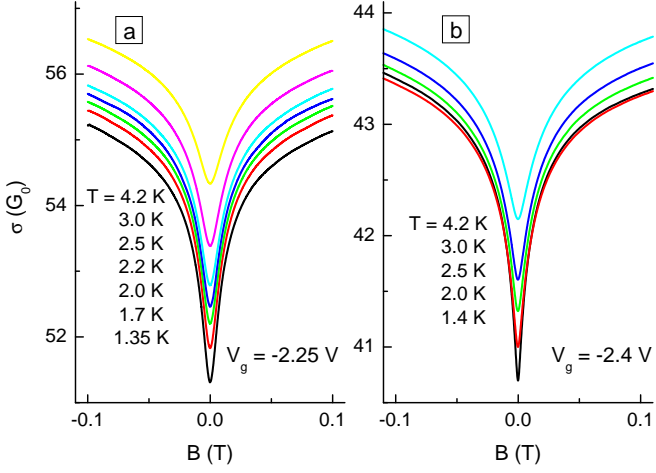}
\caption {\label{F4}  (Color online) The low-field
magnetoconductivity for different temperatures for  the pattern (a)
and unpattern (b) structure. The local conductivity is shown in
panel (a).}
\end{figure}

Let us now  inspect the low magnetic  field negative
magnetoresistivity which results from  suppression of the WL
correction. We focus our consideration on the results obtained
within the second range of the gate voltages: $V_g\simeq -(2\ldots
3)$~V. The magnetic field dependences of the local conductivity
measured in units of $G_0=e^2/(2\pi^2\hbar)\simeq 1.23\times
10^{-3}$~$\Omega^{-1}$ are presented for different temperatures in
Fig.~\ref{F4}(a). For comparison, the analogous dependences for the
unpattern structure measured at close conductivity value are shown
in Fig.~\ref{F4}(b).  For the first sight the magnetoresistance
curves in the panels are very similar. However, this impression is
wrong. The difference in magnetoresistivity shape for these
structures is more pronounced when comparing the results of data
treatment performed in a standard manner.

The shape of low-field positive magnetoconductivity
$\Delta\sigma(B)=\rho_{xx}^{-1}(B)-\rho_{xx}^{-1}(0)$ caused by
suppression of the weak localization in homogeneous 2D gas is
described by the Hikami-Larkin-Nagaoka (HLN)
expression\cite{hik,schm}
\begin{eqnarray}
\Delta\sigma(B)&=&\alpha\, G_0\,
{\cal H}\left(\frac{\tau}{\tau_\phi},\frac{B}{B_{tr}}\right), \nonumber \\
{\cal H}(x,y)& = & \psi\left(\frac{1}{2}+\frac{x}{y}\right) -
\psi\left(\frac{1}{2}+\frac{1}{y}\right)- \ln{x}, \label{eq2}
\end{eqnarray}
where $B_{tr}=\hbar/(2el^2)$ is the transport magnetic field, $\tau$
is the momentum relaxation time,  $\psi(x)$ is a digamma function,
and $\alpha$ is the prefactor, which is equal to unity in the
diffusion regime ($B<B_{tr}$, $\tau\ll\tau_\phi$) and at high
conductivity ($\sigma\gg G_0$).

\begin{figure}
\includegraphics[width=\linewidth,clip=true]{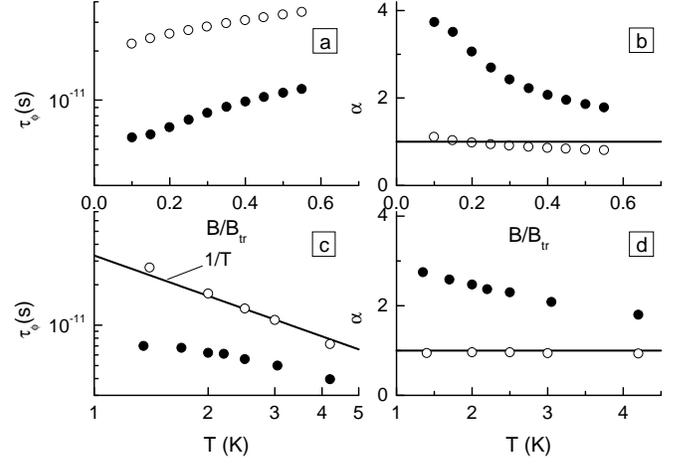}
\caption {\label{F5} The magnetic field and temperature dependences
of the fitting parameters $\tau_\phi$ and $\alpha$ for pattern
(solid symbols) and unpattern (open symbols) structures. The data in
(a) and (b) are shown for the lowest temperatures, the data in (c)
and (d) are obtained for the fitting interval from 0 up to
$0.25B_{tr}$.}
\end{figure}

We have used this expression to fit each experimental curve within
different range of magnetic field with $\alpha$ and $\tau_\phi$ as
the fitting parameters. The results are shown in Fig.~\ref{F5}. How
the fitting parameters $\tau_\phi$ and $\alpha$ depend on the range
of magnetic field  is shown in Figs.~\ref{F5}(a) and \ref{F5}(b),
whereas their temperature dependence is shown in Figs.~\ref{F5}(c)
and \ref{F5}(d). One can see that the parameters found for the
unpattern structure  behave themselves reasonable. Their values only
slightly depend on the fitting interval.\cite{footn2} The prefactor
is close to unity and practically independent of temperature. The
temperature dependence of $\tau_\phi$ is close to the theoretical
one $\tau_\phi\propto 1/T$. Thus, the HLN expression well describes
$\Delta\sigma(B)$ for the unpattern sample.

In contrast to this, the strong sensitivity of the fitting
parameters to the fitting interval takes place for the pattern
sample. Therewith the value of $\alpha$ is significantly larger than
unity and strongly dependent on the temperature. The value of the
fitting parameter $\tau_\phi$ is much less than that for the
unpattern structure and it saturates with the decreasing
temperature. All of this means that the WL correction for the
pattern structure is not described by Eq.~(\ref{eq2}) and the
determination of the phase breaking time by the standard way is
impossible.\cite{footn3}

The strong dependence of the fitting parameters on the fitting
interval of the magnetic field indicates that the role of dielectric
antidots in 2D gas is not reduced to arising of the prefactor
$\exp{(-\tau_E/\tau_\phi)}$ with $\tau_E$ as the Ehrenfest time,
which suppresses the weak localization in the 2D gas with hard discs
as scatterers\cite{aleiner9697,whit08,exper4}.

\begin{figure}
\includegraphics[width=0.55\linewidth,clip=true]{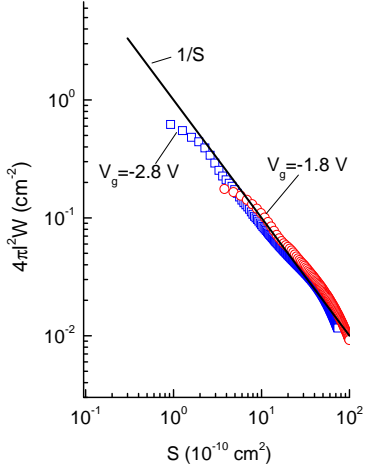}
\caption {\label{F69}  (Color online) The area distribution function
$W(S)$ obtained experimentally for the unpattern samples for two
gate voltages (symbols). Line is the diffusion limit: $4\pi l^2
W=1/S$. }
\end{figure}

Specific features of the weak localization in the pattern sample can
be understood by considering the quasiclassic interpretation of this
phenomenon. Within quasiclassic approximation the conductivity
correction is expressed through the classical quasiprobability for
an electron to return to the area of the order $\lambda_F l$
($\lambda_F=2\pi/k_F $, $k_F$ is the Fermi wave vector) around the
start point\cite{gork,chak,dyak,dmit}
\begin{equation}
\delta\sigma=-\sigma_0 \frac{\lambda_Fl}{\pi} {\cal W}, \label{eq3}
\end{equation}
where $\sigma_0= \pi k_F l G_0$ is the Drude conductivity, and
${\cal W}$ stands for the quasiprobability density of return ({\it
quasi}- means that ${\cal W}$ includes not only the classical
probability density, but the interference destruction due to an
external magnetic field and inelastic scattering processes). With
taking these effects into account  Eq.~(\ref{eq3}) can be rewritten
as follows\cite{our2}
\begin{eqnarray}\label{eq4}
    \delta\sigma(b)&=&-2\pi l^2 G_0 \int_{-\infty}^\infty dS\ \biggl\{W(S)
    \nonumber\\
    & &\exp\left[-\frac{\overline{L}(S)}{l_\phi}\right]\cos
  \left(\frac{b S}{l^2}\right)\biggr\},
\end{eqnarray}
where  $l_\phi$ is the phase breaking length connected with
$\tau_\phi$ through the Fermi velocity, $l_\phi=v_F\tau_\phi$,
$W(S)$ and $\overline{L}(S)$ are the algebraic area distribution
function of closed paths and the area dependence of the average
length of closed paths respectively (for more detail see Sec. II of
Ref.~\onlinecite{our2}). This equation shows that the shape of the
magnetoconductivity curve is determined by the statistics of the
closed paths, namely by the area distribution function $W(S)$ and by
function $\overline{L}(S)$. It is clear that the existence of
dielectric dots in 2D gas should change the statistics of closed
paths resulting in the change of  the shape of the
magnetoconductivity curve. In Ref.~\onlinecite{our3}, there is shown
how the analysis of the Fourier transform of the negative
magnetoresistance provides the information on the function $W(S)$.
The short of the matter is clear from Eq.~(\ref{eq4}). It is seen
that the Fourier transform of $\delta\sigma(B)$
\begin{displaymath}
   \Phi (S)=\frac{1}{\Phi_0}\int_{-\infty}^{\infty}dB\ \delta\sigma(B)
 \cos\left(\frac{2\pi B S}{\Phi_0}\right)
 \label{eq5}
\end{displaymath}
is equal to
\begin{equation}
  \label{eq6}
   \Phi (S)=-2\pi l^{2} G_{0} W(S)\exp \left( -\frac{ \overline{L}
(S)}{l_\phi} \right),
\end{equation}
where $\Phi_0=\pi\hbar/e$ is the elementary flux quantum. Since
$l_{\phi }$ tends to infinity when $T\to 0$, the extrapolation of
$\Phi(S,T)$ to $T=0$ should give the value of $ 2\pi l^{2}
G_{0}W(S)$. It is ideal situation. In the reality, such the approach
allows us to obtain experimentally the area distribution function of
the  trajectories which length is less than approximately
$2\,l'_\phi$, where $l'_\phi$ is the phase relaxation length at
lowest temperature of the experiment, because the contribution of
longer trajectories to the magnetoconductivity is very small.

\begin{figure}
\includegraphics[width=\linewidth,clip=true]{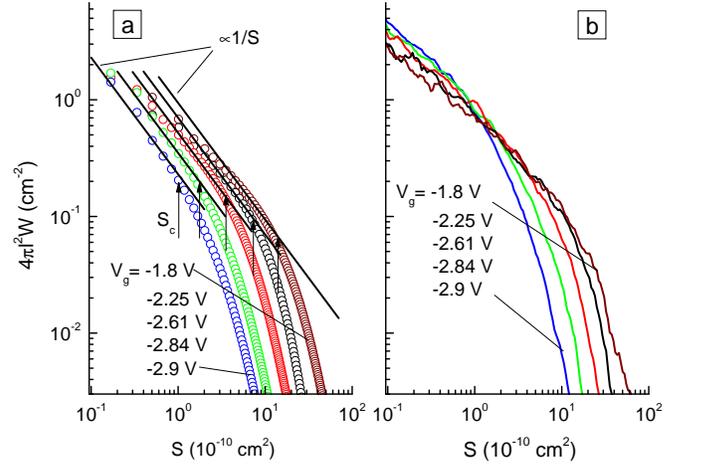}
\caption {\label{F6}  (Color online) (a) The function $4\pi l^2
W(S)$ obtained experimentally for the pattern sample for different
gate voltages. Arrows indicates where the plots deviate from
$1/S$-dependence. (b) The function $4\pi l^2W(S)$ obtained from the
simulation procedure carried out with parameters corresponding to
the gate voltages in panel (a). }
\end{figure}

The results of the data processing for the unpattern and pattern
samples are shown in Figs.~\ref{F69} and \ref{F6}(a), respectively
(note the scales in these figures are identical). It is seen that
not only the $W$-vs-$S$ dependences are drastically different for
the pattern and unpattern samples but their responses to the change
of the gate voltage as well. The area distribution function for the
unpattern sample is very close to that predicted theoretically for
2D homogeneous systems in the diffusion regime, $4\pi l^2W=S^{-1}$,
$S>l^2$ (see Refs.~\onlinecite{our2} and \onlinecite{our3}), it is
practically insensitive to the gate voltage (Fig.~\ref{F69}). As for
the pattern sample, it is seen that the curves follow near the $1/S$
dependence only at low $S<S_c$ values. At higher areas, $S>S_c$, $W$
decreases with $S$ increase much steeper than $S^{-1}$; the lower
gate voltage is the lower areas $S_c$ at which this deviation occurs
[Fig.~\ref{F6}(a)].

Before to interpret such the behavior of $W(S)$ for the pattern
sample let us clarify the meaning of this quantity for this case.
Because our system is inhomogeneous due to dielectric inclusions
under the holes in the cap layer, Eq.~(\ref{eq4}) should be
rederived. This is because the function $W(S)$ is not universal now,
it should depend on the position of starting point. If one neglects
as above the aperiodicity of the antidots array, we obtains the
following expression connecting the correction to the conductivity
of the antidots array and local correction $\delta\sigma_l$:
\begin{equation}
  \label{eq7}
 \delta\sigma(b)=\frac{\displaystyle\int\frac{dy}{[w(y)]^2}\int dx\,\delta\sigma_l(x,y,b)}{\left[\displaystyle\int
\frac{dy}{\displaystyle
 w(y)}\right]^2}.
 \end{equation}
Here, the integration runs over the intervals given by the border of
conducting area, $w(y)$ is the $y$ dependence of width of conducting
area, and the quantity $\sigma_l(x,y,b)$ is given by the expression
analogous to Eq.~(\ref{eq4}):
\begin{eqnarray}\label{eq8}
    \delta\sigma_l(x,y,b)&=&-2\pi l^2 G_0 \int_{-\infty}^\infty dS\ \biggl\{\mathfrak{W}(x,y,S)
    \nonumber\\
    & &\exp\left[-\frac{\overline{L}(x,y,S)}{l_\phi}\right]\cos
  \left(\frac{b S}{l^2}\right)\biggr\},
\end{eqnarray}
in which $\mathfrak{W}(x,y,S)$ is determined in such a way that
$\mathfrak{W}(x,y,S)dS$ gives the density probability of return to
the starting point with the coordinates $(x,y)$ with the enclosed
algebraic area in the interval $(S,S+dS)$.

Thus, the above described procedure, which for the homogeneous 2D
gas allows us to obtain the area distribution function of closed
paths, being applied to the data obtained for the pattern sample
gives the {\it effective} area distribution function:
\begin{equation}
  \label{eq9}
 W(S)=\frac{\displaystyle\int\frac{dy}{[w(y)]^2}\int \mathfrak{W}(x,y,S)\,dx}{\left[\displaystyle\int
\frac{dy}{\displaystyle
 w(y)}\right]^2}.
 \end{equation}

Now we are in position to discuss the behavior of $W(S)$ for the
pattern sample [Fig.~\ref{F6}(a)]. In order to understand  main
features we have performed  the computer simulation of a particle
motion over 2D plane with scatterers. The details can be found in
Refs.~\onlinecite{our2,our4,our5}, below is outline only and
important features. The 2D plane is represented as a lattice with
scatterers of two types placed in a part of lattice site. The
scatterers of the first type with isotropic differential
cross-section correspond to ionized impurity. The scatterers of the
second type are hard discs with specular reflection from the
boundaries. Particle motion is forbidden within the disks. They
correspond to the areas of 2D gas under the holes. A particle is
launched from some point with $x,y$ as  coordinates, then it moves
with a constant velocity along straight lines, which happen to be
terminated by collisions with the scatterers. After collision it
changes the motion direction. If the particle passes near the
starting point at the distance less than some prescribed value
$a/2\ll l$, the path is perceived as being closed. Its length and
enclosed algebraic area are calculated and kept in memory. The
particle walks over the plane until the path traversed is longer
than $2l'_\phi$, where $l'_\phi$ is the phase relaxation time
obtained at lowest temperature on the unpattern sample for the same
$V_g$ value. As mentioned above namely the statistics of such closed
paths can be reliably obtained from the weak localization
experiments. When the path becomes longer than this value another
particle is launched and all is repeated. After large number of
launches from the starting point with given $x,y$ coordinates
$\mathfrak{W}(x,y,S)$  is calculated as
\begin{equation}
 \mathfrak{W}(x,y,S)=\frac{n_S}{N\, l\, a\, \Delta S},
 \label{eq20}
 \end{equation}
where $N$ is the number of starts from this point, $n_S$ is the
number of returns along the trajectory  with enclosed area in the
interval $(S,S+\Delta S)$. Launching the particle from different
starting points, we are able to calculate numerically the function
$W(S)$ using Eq.~(\ref{eq9}) in the discrete form.

Shown in Fig.~\ref{F6}(b) are the results of simulation procedure
carried out with the parameters corresponding to the gate voltages
from Fig.~\ref{F6}(a). It is clearly seen that the results of
computer simulation are in qualitative agreement with those of real
experiment.

\begin{figure}
\includegraphics[width=0.9\linewidth,clip=true]{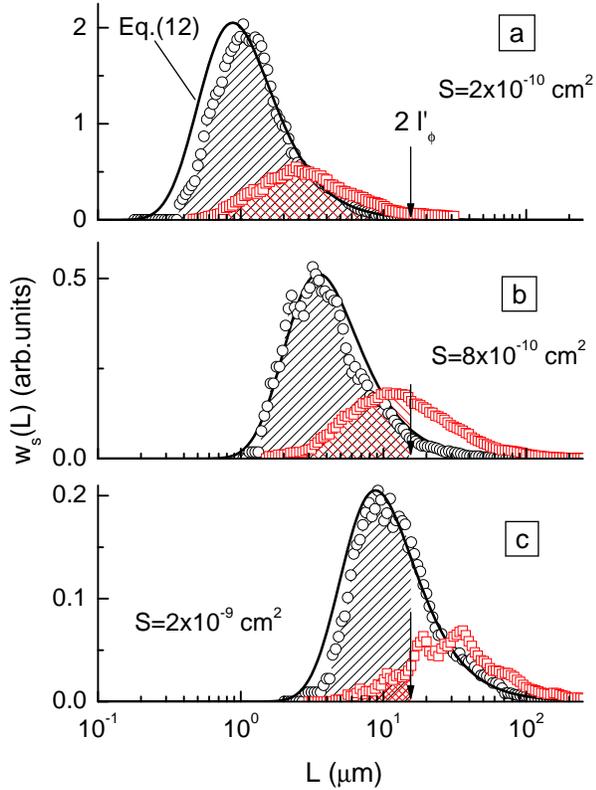}
\caption {\label{F61}  (Color online) The length distribution
functions of closed paths with different area enclosed obtained from
the simulation procedure with parameters corresponding to  the
unpattern (circles) and pattern (squares) samples  for
$V_g=-2.84$~V. Lines are calculated  from Eq.~(\ref{eq30}) with
corresponding $S$-values and $l=0.06$~$\mu$m. Hatched areas are the
$W$ value for paths with $L\leq 2\,l'_\phi$. }
\end{figure}

Analysis of the simulation results gives an insight into why the
experimental $W$-vs-$S$ dependence is steeper for the pattern
sample. This can be done if one considers how the paths with the
given area enclosed are distributed over the length. This
distribution is characterized by the function $w_S(L)$ determined in
such a way that $w_S(L)dS$ gives the density probability of return
along a trajectory with the length $L$ and area in the interval
$(S,S+dS)$. In Fig.~\ref{F61} we show the functions $w_S(L)$
obtained from the simulation procedure for the pattern\cite{footn4}
and unpattern samples (all the paths including those for which
$L>2\,l'_\phi$ are taken into account here). Qualitatively, the
functions are similar. They have a peak which position characterizes
the typical length of closed paths. In both cases, the larger area
enclosed the longer trajectories. The $w_S$-vs-$L$ plot for the
unpattern sample is well described by the expression
\begin{equation}
 w_S(L)=\frac{1}{(2\,l\,L)^2}\cosh^{-2}\left(\frac{\pi S}{l\,L}\right)
 \label{eq30}
 \end{equation}
obtained analytically  within the diffusion approximation for the
ordinary 2D system\cite{our2} that justifies the validity of
simulation procedure. An important point is that the closed paths in
the pattern sample are significantly longer than the paths in the
unpattern sample  with the same area enclosed. If, in correspondence
with the experimental situation,  one restricts the consideration by
short trajectories, $L\lesssim 2\,l'_\phi$, we obtain
\begin{equation}
 W(S)=\int_0^{2\,l'_\phi} w_S(L)\frac{dL}{l}.
 \label{eq35}
\end{equation}
approximately the same for the both samples when $S$ is relatively
small [Figs.~\ref{F61}(a)], and much less in the pattern sample when
$S$ is sufficiently large [Fig.~\ref{F61}(c)]. Thus, the different
length distribution of closed paths is the reason of the different
behavior of the area distribution function obtained experimentally
for the pattern and unpattern samples.

Let us return to the magnetoconductivity caused by suppression of
the interference  quantum correction. The simulation procedure
allows us to calculate $\delta\sigma(b)$ for the model system. This
can be done with the use of Eq.~(\ref{eq7}) in the discrete form and
of $\delta\sigma_l(x,y,b)$ calculated as
\begin{equation}
  \label{eq11}
 \delta\sigma_l(x_i,y_i,b)=\frac{2\pi l G_0}{d\,N_i}\sum_k
 \cos\left(\frac{bS_i^k}{l^2}\right)\exp{\left(-\frac{l_i^k}{l_\phi}\right)},
 \end{equation}
where summation runs over all closed trajectories among a total
number of trajectories $N_i$ starting from the point with $x_i$ and
$y_i$ as coordinates, $S_i^k$ and $l_i^k$ stand for algebraic area
and length respectively of the $k$-th trajectory.

\begin{figure}
\includegraphics[width=\linewidth,clip=true]{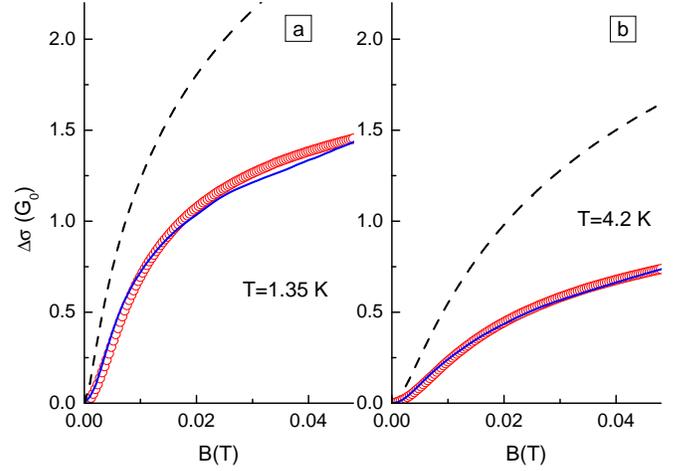}
\caption {\label{F7} (Color online) The low-field magnetoconductance
for the pattern sample for $T=1.35$~K (a) and $4.2$~K(b),
$V_g=-2.25$~V. Symbols are the measured dependences. Solid and
dashed lines are the simulation results and Eq.~(\ref{eq2}) obtained
with $\tau_\phi=1.9\times 10^{-11}$~s (a) and $\tau_\phi=6.8\times
10^{-12}$~s (b). }
\end{figure}

In Fig.~\ref{F7} we compare the simulation and experimental results.
Symbols are the experimental plot obtained for the pattern sample at
$V_g=-2.25$~V. Solid lines are the simulated magnetic field
dependences of $\Delta\sigma=\delta\sigma(B)-\delta\sigma(0)$. The
values of $\tau_\phi=1.9\times 10^{-11}$~s for $T=1.35$~K and
$\tau_\phi=6.8\times 10^{-12}$~s for $T=4.2$~K correspond to the
best accordance with the experimental data. It is significant that
they are close to that obtained for the unpattern sample [see
Fig.~\ref{F5}(c)]. If one calculates $\Delta\sigma(B)$ from
Eq.~(\ref{eq2}) with the same parameters and $\alpha=1$ we obtain
drastic disagreement with the experimental magnetoconductance (see
dashed lines Fig.~\ref{F7}).

Thus, the simulation approach allows us to understand qualitatively
the main features of the low-field magnetoconductivity in the
pattern samples, which are determined by peculiarities of statistics
of closed paths.

It would serve no purpose to make more detailed comparison between
the experimental and simulation results because our model is rather
crude. So, we supposed that the areas forbidden for classical motion
had the form of hard disks. As seen from Fig.~\ref{F0}(b) it is not
exactly. Moreover, it was suggested that the mean free path and
Fermi velocity are constant over all the conducting  area that
should not be fulfilled in the real structures.

In summary, we have studied the weak localization in the 2D electron
gas with potential electrostatic relief forming the insulating
antidots array. It has been shown that the use of the standard
procedure for the obtaining of the phase relaxation time from the
shape of the magnetoconductivity curve is inadequate when it is
applied to the antidots array structure in the diffusion regime. To
understand the main features  of the weak localization in these
systems the alternative approach based on the analysis of the
statistics of closed paths has been used. We have shown that the
main peculiarities of the WL phenomenon in the pattern structure are
due to the specific  statistics of closed paths. The paths of actual
areas in system with antidots are characterized by significantly
larger lengths as compared with the usual 2D systems.

\subsection*{Acknowledgments}
We are grateful to I.~V. Gornyi for very useful discussions and
S.~V.~Dubonos for sample fabrication. This work was supported in
part by the RFBR (Grant Nos. 06-02-16292, 07-02-00528, and
08-02-00662), the CRDF (Grant No. Y3-P-05-16).


\begin{thebibliography}{}
\bibitem{footn1} Another way of
description of the positive magnetoresistance is considering of
interplay  between short and long range disorder in the magnetic
field.\cite{Theory1}
\bibitem{footn4} For the pattern sample, the effective length
distribution function  defined by analogy with Eq.~(\ref{eq9}) is
presented.
\end{thebibliography}
\end{document}